# Investigation of muon flux anisotropy during CME

I.I. Astapov[1], N.S. Barbashina[1], V.V. Borog[1], I.S. Veselovsky[1,2,3],
N.V. Osetrova[1], A.A. Petrukhin[1], V.V. Shutenko[1]

[1]*National Research Nuclear University MEPhI (Moscow Engineering Physics Institute), 115409 Moscow, Russia*

[2]*Skobeltsyn Institute of Nuclear Physics, Lomonosov Moscow State University (MSU), Moscow 119991, Russia*

[3]*Space Research Institute (IKI), Russian Academy of Sciences, Moscow 117997, Russia*

According to CACTus catalog, during periods of a high solar activity every day up to tens of coronal mass ejections are observed. Such ejections have an impact on the flux of cosmic rays that permeate the space around us. Unlike most ground cosmic ray detectors, muon hodoscope URAGAN (MEPhI) allows to investigate not only the integrated counting rate of registered particles, but also the spatial and angular characteristics of the muon flux at ground level. This approach to particle detection allows fixing changes in the flux of cosmic rays not only for geoeffective CMEs, but also for the ejections, the front of which is directed to the opposite side of the Sun. The results of the study of different types of CMEs at different stages of the solar activity from 2008 to 2015 are presented.

## 1. INTRODUCTION

Cosmic rays represent a powerful instrument for the study of a variety of dynamic processes in the interplanetary space, including coronal mass ejections. The disturbances caused by the solar activity have a direct impact on the fluxes of primary and, therefore, secondary cosmic rays. In recent years, a new direction in investigation of the processes in the interplanetary space – muon diagnostics [1] – becomes more and more popular. Muon diagnostics is based on the analysis of the muon flux registered by the wide-aperture coordinate detectors – muon hodoscopes – which allow real-time reconstruction of tracks of all muons coming from the celestial hemisphere [2, 3].

The coronal mass ejections (CMEs) occurred during the period from 2008 to 2015 and registered by coronagraphs LASCO C2/C3 are considered. Event selection is performed using the CACTUS data base [4].

In this paper, the results of the analysis of data of the muon hodoscope URAGAN – the part of the Unique Scientific Facility 'Experimental complex NEVOD' (MEPhI, Moscow) – which allows to explore not only the integral counting rate of the flux of registered muons, but also its local anisotropy, are presented.

## 2. MUON HODOSCOPE URAGAN AND EXPERIMENTAL DATA

Muon hodoscope (MH) URAGAN [2] (55.7°N, 37.7°E, 173 m above sea level) is the coordinate detector that allows to investigate variations of the muon flux angular distribution on the Earth's surface. Muons retain the direction of the primary particles motion, which allows to study changes in primary cosmic ray flux in the interplanetary space. URAGAN consists of four independent supermodules (SM). Each SM is assembled of eight layers of gas-discharge chambers (streamer tubes) equipped with two-coordinate system of external readout strips which provides a high spatial and angular accuracy of muon track detection (correspondingly, 1 cm and 1°) in a wide range of zenith (0°-80°) and azimuthal (0°-360°) angles in a real-time mode. Experimental data accumulated by the supermodules of the URAGAN setup equipped with a multi-channel measuring system are binary files that contain information about one-minute frames, which are formed by the control program during the setup operation.

The supermodule response represents the information about triggered strips in each of two projection planes XZ and YZ. Supermodules are oriented such that the X and Y axes are turned clockwise by 35° relative to the South and East geographical axes.

For the study of two-dimensional variations of muon flux registered by the muon hodoscope URAGAN, a local anisotropy vector $\vec{A}$ [5] which is the sum of the unit vectors of particle tracks normalized by the total number of tracks is used (see figure 1).

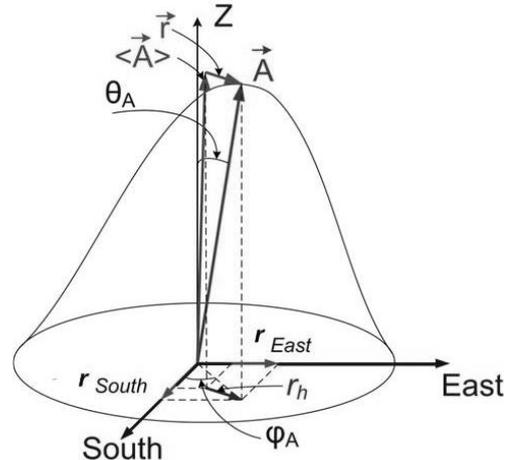

FIG. 1: Vector of local anisotropy of the muon flux.

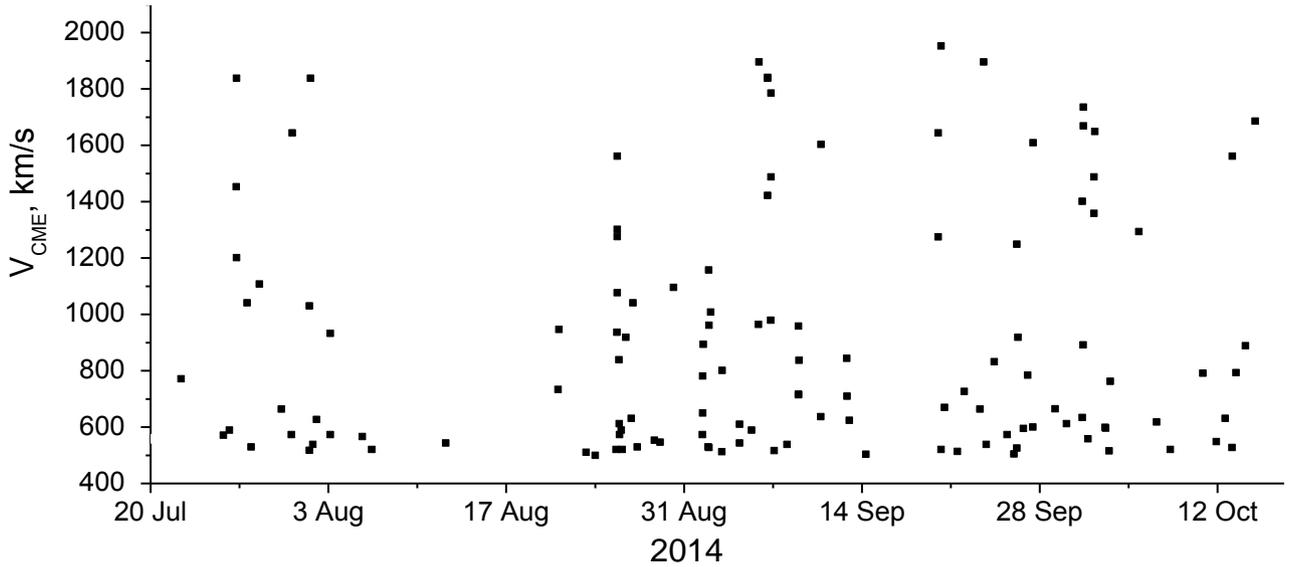

FIG. 2: Coronal mass ejections during the period from 20 July to 20 October, 2014. The points mark the beginning of time of the CME and the corresponding average speed.

Local anisotropy vector $\vec{A}$ indicates the average arrival direction of muons which is close to the vertical. To study its deviations from the mean value $<\vec{A}>$, the relative anisotropy vector which represents the difference between the current vector and the average anisotropy vector calculated over a long period of time is used: $\vec{r} = \vec{A} - <\vec{A}>$. For the analysis of muon flux variations, the horizontal projection of the relative anisotropy vector $r_h$ which characterizes the 'side impact' on the muon flux angular distribution is used: $r_h = \sqrt{r_x^2 + r_y^2}$.

Its value depends on the state of the heliosphere and can be of the order of $10^{-4}$–$10^{-3}$. To estimate the statistical significance of the observed deviations, the experimental data of 2009 (minimum of solar activity) were used. For 2009 the mean value is $2.45\times10^{-4}$ and rms-deviation of $r_h$ is $1.28\times10^{-4}$. Taking into account that distribution of $r_h$ is close to the Rayleigh's one, for values more than $r_h/\sigma_{rh} > 5$ the probability of random deviation is less than 1%.

### 3. CME SELECTION

The CACTUS data base is formed automatically according to the data of the LASCO C2/C3 coronagraphs and contains information about maximal, average and minimal velocity of ejections, time of CME occurrence, principal angle (direction) and angular width. The events with average velocity of 500 km/h were considered. In figure 2, the beginnings and the corresponding average velocities of the CMEs occurred during the period from July 20 to October 20, 2014 are shown. In this period, 129 ejections with velocity of more than 500 km/s were observed. In total, 2237 events were observed from 2008 to 2015.

Figure 3 shows distribution of the $\Delta t_{0\_CME}$ parameter which characterizes the time interval between the beginnings of some CME and the following one.

It is seen that for 80% of events the time interval between the beginnings of CMEs is less than 24 hours. The influence of separate events occurred during the time interval less than 24 hours on the cosmic ray flux cannot be accounted. So, such events were grouped. If the time between nearby ejections is less than 12 hours, these CMEs are combined into one group. All selected CMEs were combined into 780 groups. These groups were classified by the number of CMEs: weak (2-4 CMEs), average (5-10 events) and powerful (more than 10 events). Figure 4 shows the distribution of CMEs by years from 2008 to 2015 before and after combining into groups.

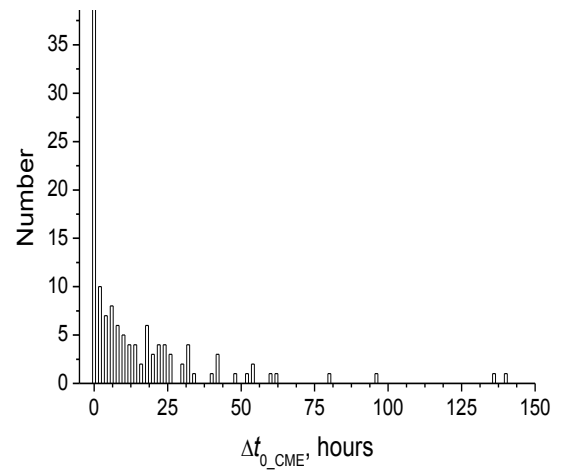

FIG.3: Distribution of the $\Delta t_{0\_CME}$ parameter.

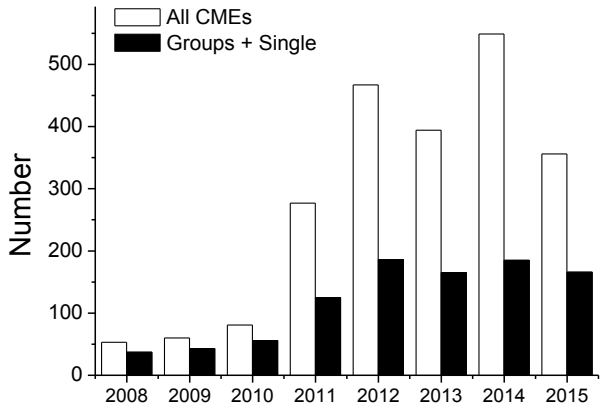

FIG. 4: Distribution of CMEs by years from 2008 to 2015 before and after combining into groups.

## 4. RESULTS AND DISCUSSIONS

Figure 5 shows an example of the muon hodoscope URAGAN response to the coronal mass ejections which occurred on July 19-20, 2015. In this period, 6 CMEs were observed. Maximal velocity in this group was about 1300 km/s. The average velocity of the group was ~ 800 km/s. The bottom plot shows the data on muon flux local anisotropy ($r_h/\sigma_{rh}$ parameter) according to the MH URAGAN data. The mean value $<r_h/\sigma_{rh}>_{2008-2015}$ for the period from 2008 to 2015 is 2.93 (standard deviation SD = 1.81).

It is seen that the response to the group was observed on July 21 (maximal $r_h/\sigma_{rh}$ value was 11.8) an a exceeded the average value by 4 times. The effect of ejections on the muon flux maintained for the following 24 hours.

The detailed analysis was carried out for all events. Figure 6 shows the results of studying the response of muon hodoscope to different types of events in the period 2008-2015. The distributions of amount of studied CMEs and the number of the MH URAGAN responses to these CMEs by years of observation are given in the figure.

The muon hodoscope response to powerful groups (more than 10 CMEs) and to average groups (5-10 CMEs) appears in about (75 – 90) % of events depending on the year of observation. In years of high solar activity (2012 – 2015) the response of the muon hodoscope URAGAN to weak groups (2-4 CMEs) is observed in about (55 – 70) % of events. In years of low solar activity (2008-2011) the response was observed an average in ~ 30 % of events. The response of muon hodoscope to single CMEs is quite weak. In years of high solar activity the muon hodoscope response to such CMEs was observed in ~30 % of events, and in years of low activity only in ~20 % of events.

Thus, the studying of muon flux local anisotropy allows to select powerful ejections which in their turn can affect the Earth's magnetic field. All selected groups of CMEs were analyzed in a similar way.

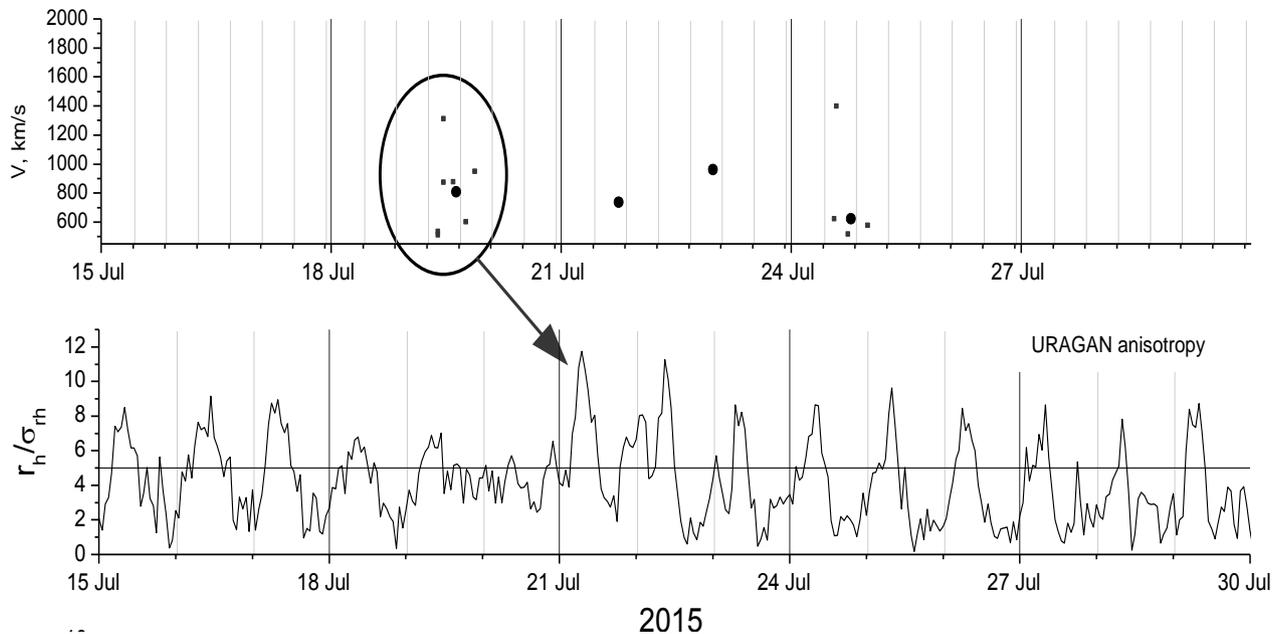

FIG. 5: Example of the muon hodoscope URAGAN response to the group of CMEs occurred on July 19-20, 2015.

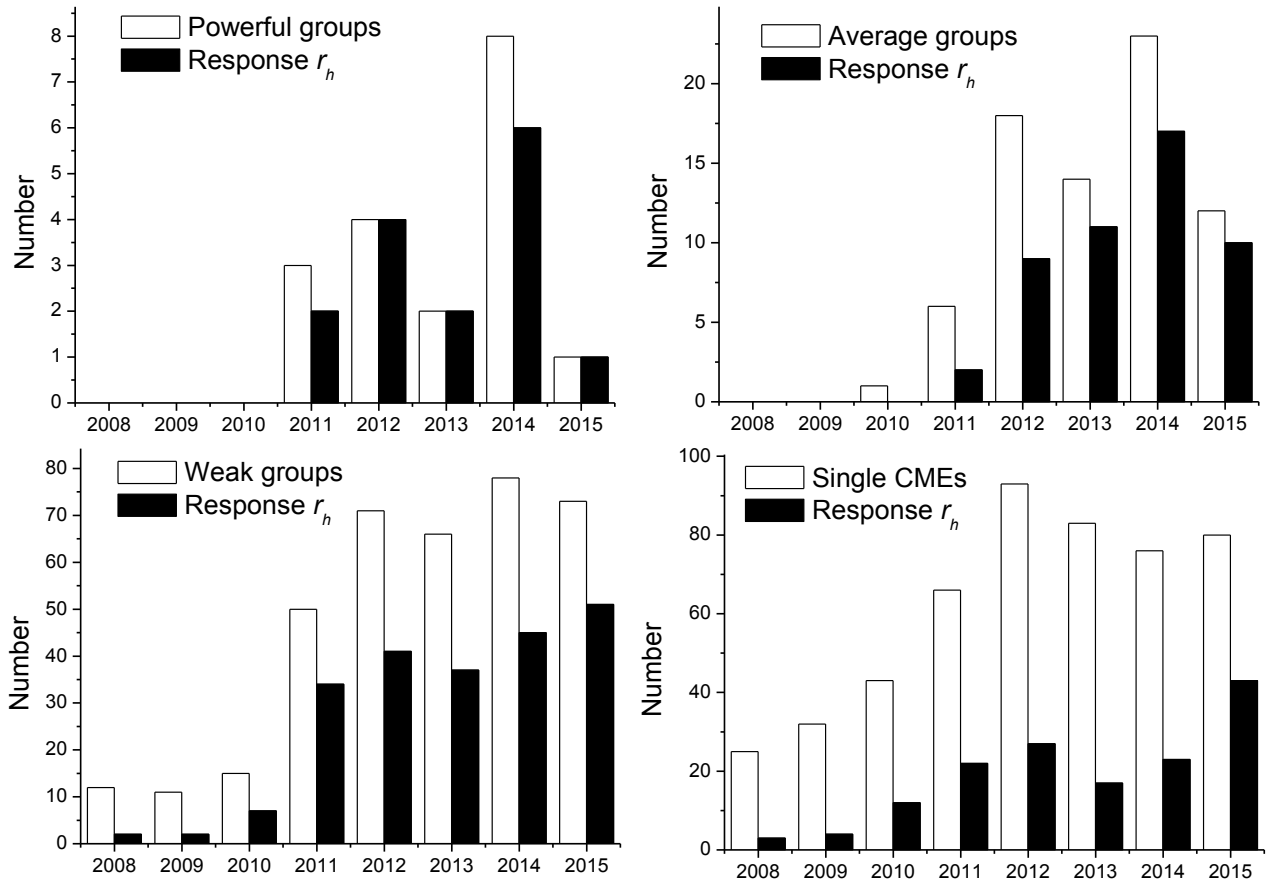

FIG. 6: Distributions of CMEs and responses of the muon hodoscope URAGAN to these CMEs in the period 2008-2015.

## 5. CONCLUSIONS

Registration of muons in the hodoscopic mode allows studying the cosmic ray flux local anisotropy sensitive to the disturbance in the heliosphere using a single setup. The described approaches to CME separation illustrate wide opportunities of application of this method for reliable identification of geoeffective disturbances in the near-Earth space.

## Acknowledgments

The work was performed at the Unique Scientific Facility "Experimental complex NEVOD" with the support of the MEPhI Academic Excellence Project (contract № 02.a03.21.0005, 27.08.2013).